\documentclass[aps,floats,floatfix,superscriptaddress]{revtex4}
\usepackage{amsfonts,amsmath,amssymb,hyperref}
\usepackage{filecontents}
\usepackage[pdftex]{graphicx}
\usepackage{color}
\usepackage{times}

\begin{document}
\title{Long-Range Correlations and Memory in the Dynamics of Internet Interdomain Routing}

\author{Maksim Kitsak}
\affiliation{Department of Physics, Northeastern
University, 110 Forsyth Street, 111 Dana Research Center, Boston, MA 02115, USA.}
\affiliation{Center for Cancer Systems Biology and Department of Cancer Biology, Dana-Farber
Cancer Institute and Department of Genetics, Harvard Medical School, 450 Brookline Ave., 02215
Boston, USA}

\author{Ahmed Elmokashfi}
\affiliation{Simula Research Lab, Oslo, Norway}

\author{Shlomo Havlin}
\affiliation{Department of Physics, Bar-Ilan University, Ramat Gan 52900, Israel}

\author{Dmitri Krioukov}
\affiliation{Department of Physics, Department of Mathematics, Department of Electrical\&Computer Engineering,
Northeastern University, 110 Forsyth Street, 111 Dana Research Center, Boston, MA 02115, USA.}

\maketitle

\section*{Abstract}

 Data transfer is one of the main functions of the Internet. The Internet consists of a large number of interconnected subnetworks or domains, known as Autonomous Systems (ASes). Due to privacy and other reasons the information about what route to use to reach devices within other ASes is not readily available to any given AS. The Border Gateway Protocol  (BGP)  is responsible for discovering and distributing this reachability information to all ASes. Since the topology of the  Internet is highly dynamic, all ASes constantly exchange and update this reachability information in small chunks, known as routing control packets or BGP updates. In the view of the quick growth of the Internet there are significant concerns with the scalability of the BGP updates and the efficiency of the BGP routing in general. Motivated by these issues  we conduct a systematic time series analysis of BGP update rates. We find that BGP update time series are extremely volatile, exhibit long-term correlations and memory effects, similar to seismic time series, or temperature and stock market price fluctuations. The presented statistical characterization of BGP update dynamics could serve as a basis
for validation of existing and developing better models of Internet interdomain routing.

\section{Introduction}

On large scale, the Internet is a global system of approximately $40,000$ interlinked computer
networks connecting billions of users and devices worldwide~\cite{CIDRReport}. These networks are
called Autonomous Systems (ASes). ASes vary in size and function: they can  be (i) Internet Service
and/or Transit Providers ($AT\&T$), (ii) Content Providers (Google), (iii) Enterprises (Harvard
University), and (iv) Non-profit organizations~\cite{rfc1930}.  Devices inside ASes are identified
via unique Internet Protocol (IP) addresses, which are 32- or 128-bit numerical labels that act
both as identifiers and locators of devices. An IP address is divided into two sections, a network
section and a host section. The network section, which is known as IP prefix, identifies a group of
hosts, while the host section identifies a particular device. An AS can include a number of IP
prefixes.

Each AS is administrated by a single entity, but a single organization may own and operate several
ASes. ASes connect to each other via contractual agreements that govern the flow of data between
and through them. This interconnection of ASes shapes the AS-level topology of the Internet, which
facilitates connectivity between any pair of ASes and thus any pair of devices connected to the
Internet (Fig.~$1${\bf a}).

The information about how to reach devices within other ASes is not readily available to them. The
exchange of this information is handled by specialized networked computers called routers.
Performing routing requires signaling reachability information, comparing different possibilities,
and maintaining a state that describes how to reach different IP prefixes. The Border Gateway
Protocol (BGP)~\cite{rfc4271} is the globally deployed routing protocol that accomplishes this
task. The BGP protocol can be summarized as follows. ASes advertise their IP prefixes to their
neighbor ASes through BGP update messages. At each AS incoming BGP updates are processed by the BGP
router and the resulting reachability information is then stored in routing tables.

The Internet is a dynamic system where participating networks and links between them do often
experience configuration changes, failures, and restorations. BGP protocol reacts to changes in the
Internet connectivity incrementally: BGP routers send update messages to their neighbor BGP
routers. BGP update messages do not carry the information on the whole Internet connectivity state.
Instead, they carry only the information concerning the affected IP prefixes. Hence, to keep a
consistent view of the network and, consequently, to be able to communicate with other networks, a
BGP router must process incoming BGP updates in a timely manner and  update its routing table
accordingly (Fig.~$1${\bf b, c}).

Current version of the BGP routing protocol was introduced in 1994. Since then, the deployment of
the BGP routing protocol has sustained tremendous growth and it is arguably one of the main
technological reasons behind the success of the Internet.

\begin{figure}
\centering
\includegraphics[width=17.0cm,angle=0]{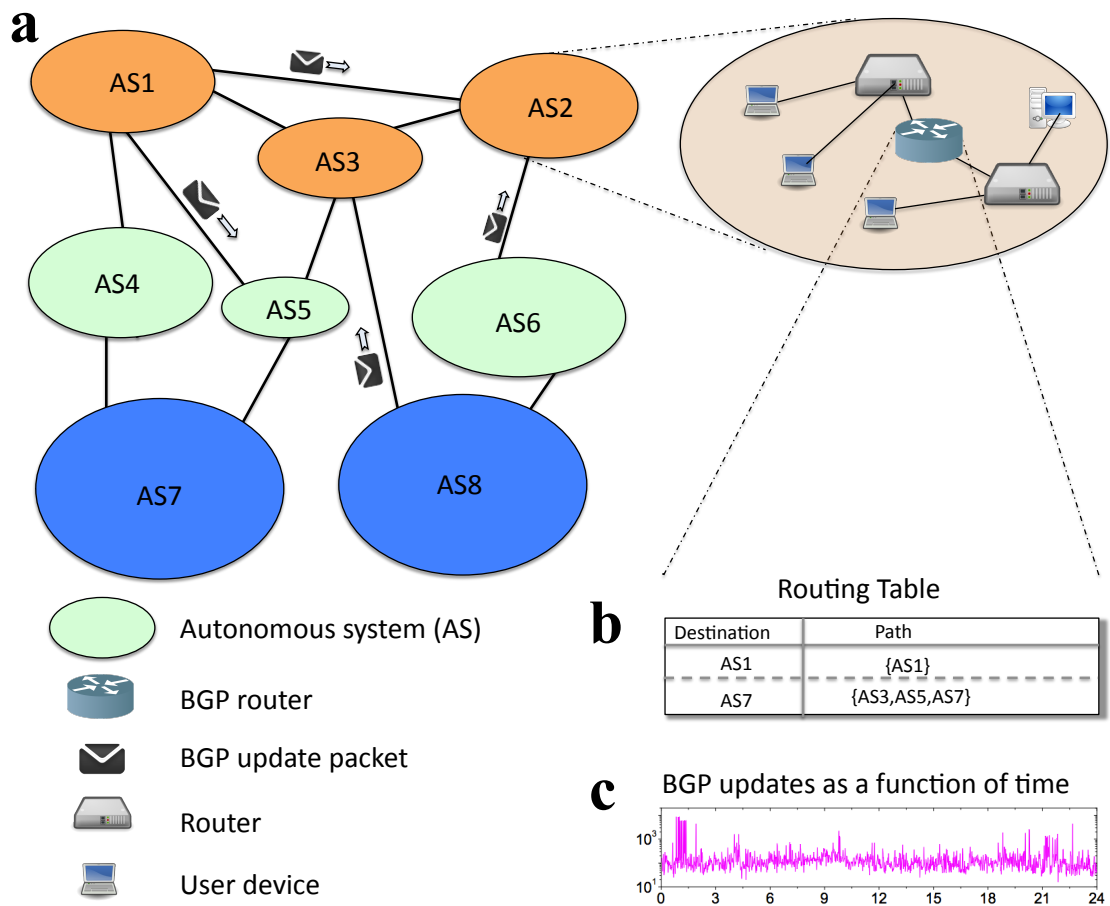}
 \caption{{\bf The Internet and BGP routing.} {\bf a}, On large scale the Internet is the
product of interconnectivity among a large number of ASes (shown with ovals). In order to perform
data transfer, ASes need to exchange the reachability information through BGP update messages. {\bf
b, c} BGP updates are processed by the BGP routers. {\bf b}, The reachability information is stored
in the routing table. {\bf c}, Typical dynamics in the number of updates received by the BGP
router.}
 \label{fig:sketch}
\end{figure}

Nevertheless, there are two major concerns related to the fast rate of the Internet growth. On one hand, Internet
growth implies the growth in the number of destinations for the BGP routing and, thus, results in
the growth of routing table sizes. On the other hand, the growth of the Internet also leads to the
growth in the number of BGP updates needed to maintain BGP routing~\cite{iabreport}. Both factors
are important, especially for routers at the core of the Internet. The growing size of routing
tables requires increasingly larger and faster memory. At the same time, growing routing table
sizes do not necessarily slow down data forwarding as long as address lookups are performed using
high speed memories and constant-time matching algorithms~\cite{Varghese:2004}. Increasingly large
amounts of BGP updates, on the other hand, is a more serious concern because processing BGP updates
can be computationally heavy (updating routing state, generating more updates, checking
import/export filters), and can trigger wide-scale instabilities~\cite{LWang:2001}.

Recent studies of BGP scalability range from measurements assessing the extent of the
concern~\cite{elmokashfi2012bgp,cittadini2010evolution}  to studies suggesting radically new
routing architectures~\cite{Farinacci:2012,Boguna:2010}. Elmokashfi et al.~\cite{elmokashfi2012bgp}
analyzed the dynamics of BGP updates in four networks at the backbone of the Internet over a period
of seven years and eight months. They have shown that  on average the level of BGP updates is
increasing, but not at an alarming rate: it was shown to grow at rates similar to the growth in the
number of ASes. However, they have also illustrated that the dynamics of BGP updates is highly
volatile even at large time scales, with peak rates exceeding the daily averages by several orders
of magnitude.

The complexity of the inter-AS routing system makes it difficult to isolate different factors
behind these fluctuations~\cite{Caesar:2003,feldmann2004locating}. An approach alternative to
inferring this factors directly is to build a realistic model for the dynamics of BGP updates. To
this end, one needs an in-depth statistical characterization of fluctuations in BGP update time
series, which is the subject of this work.

We aim at improving our understanding of these fluctuations, which can help in validating existing
models~\cite{Valler:2011} and in developing better ones. To study the statistical properties of BGP
updates, we use historical BGP update logs spanning a period of 8.5 years, collected by the
RouteViews project~\cite{routeviews} from the BGP routers of four ASes ($AT\&T$, $NTT$, $IIJ$, and
$Tinet$). Throughout the manuscript we refer to these routers as monitors. A BGP update log is the
time series of BGP updates arriving at the monitor recorded in $1$ second intervals. The four ASes
analyzed in this work are among the largest Internet Service Providers (ISPs). Therefore, their corresponding BGP update traffic is a
reflection of BGP dynamics taking place in the core of the Internet, where the BGP update
volatility is believed to reach maximum rates. (Detailed information on data collection and
pre-processing can be found in Appendix~\ref{sec:app:bgp}).

To put our study in a broader context we wish to note that many natural and economic systems have also been found to exhibit extreme fluctuations. Examples include DNA sequences~\cite{peng1992long} and heartbeat intervals~\cite{peng1993long}, climate variability~\cite{PhysRevLett.81.729,eichner03}, earthquakes~\cite{lennartz08,bunde2002science},  stock markets~\cite{yamasaki05,preis2010complex,preis2012quantifying}, and languages~\cite{gao2012culturomics,petersen2012languages}.

\section{Results}
\label{sec:results}

\begin{figure*}
\includegraphics[width=16.0cm,angle=0]{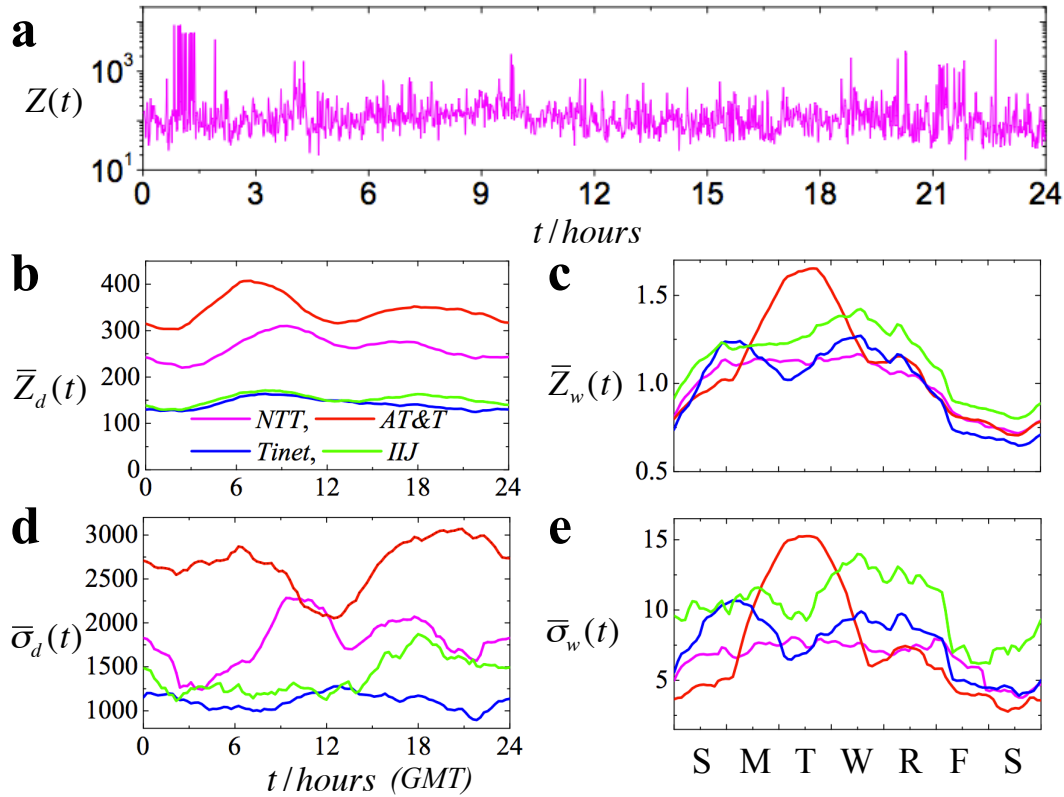}
 \caption{{\bf Time series of BGP updates.} {\bf a,} The number of updates received by the $NTT$
monitor on May 28th,
  2010, from $00.00$ to $24.00$ GMT.
 {\bf b}, the intra-day pattern, $\overline Z_{d}(t)$, and, {\bf d}, the standard
 deviation from the intra-day pattern, $\overline \sigma_{d}(t)$, of BGP updates measured for $NTT$, $IIJ$, $Tinet$,
 and $AT\&T$ monitors.
  {\bf c}, the intra-week pattern, $\overline Z_{w}(t)$, and, {\bf e}, the standard
 deviation from the intra-week pattern, $\overline \sigma_{w}(t)$, of BGP updates measured
 for $NTT$, $IIJ$, $Tinet$,
 and $AT\&T$ monitors.}
 \label{fig:fig1}
\end{figure*}

First we highlight the volatility of BGP updates series by reproducing the results of previous
works~\cite{elmokashfi2012bgp}. We plot the average rate of BGP updates received by the $NTT$
monitor on May 28th, 2010, from $00\colon00$ to $24\colon00$ Greenwich Mean Time (GMT). As seen
from Fig.~$2${\bf a}, in $1$ minute interval  the $NTT$ monitor receives on the average
several hundred updates, while extreme fluctuations occasionally produce $10^{4}$ updates per
minute. BGP updates are largely driven by two sources: spontaneous BGP events and maintenance
sessions. The former consist of  mostly  spontaneous updates, such as
misconfigurations, duplicate announcements and special events. Maintenance sessions, on the other
hand, are periodic by nature and happen at certain times of the day on particular days of the week.

In order to separate the two sources of fluctuations we calculate the intra-day and intra-week
patterns for the BGP update time series. The intra-day pattern, $\overline{Z}_{d}$, is then defined as the
 number of events taking place at a specific time of the day, $t_{day}$, averaged throughout the observation period:
\begin{equation}
\overline{Z}_{d}(t_{day}) = {1 \over N_{d}}\sum_{i=1}^{N_{d}} Z^{i}(t_{day}),
\end{equation}
where $N_{d}$ is the total number of days in the observation period, and  $Z^{i}(t_{day})$   is the
number of events at day $i$ at $t_{day}$.  The intra-week pattern $\overline{Z}_{w}(t_{week})$ is
defined in a similar way after first normalizing the time series with the intra-day pattern.
\begin{eqnarray}
\tilde{Z}(t) &\equiv& {Z(t) \over \overline{Z}_{d}\left(t_{day}\left(t\right)\right)},\\
\overline{Z}_{w}(t_{week}) &=& {1 \over N_{w}}\sum_{i=1}^{N_{w}} \tilde{Z}^{i}(t_{week})
\end{eqnarray}
Here $N_{w}$ is the number of weeks in the observational period and $\tilde{Z}^{i}(t_{week})$ is
the normalized number of events at week $i$ at time of the week
 $t_{week}$ (see Methods for details).

As seen from Fig.~$2${\bf b}, the intraday BGP update patterns reach maximum values in
the interval from approximately $06\colon00$ to $10\colon 00$ GMT, which is typical time for
scheduling maintenance tasks~\cite{FCDB07}. The intraweek patterns, in their turn, are
characterized by higher values during weekdays and smaller values during weekends. (see
Fig.~$2${\bf c}). We note that the standard deviations of the intraday and intraweek
patterns, $\overline{\sigma}_{d}(t)$ and $\overline{\sigma}_{w}(t)$, tend to exceed the
corresponding average values of the intra-day and the intra-week patterns by an order of magnitude,
which is consistent with the extreme burstiness of the BGP updates~(Fig.$2${\bf d} and
Fig.~$2${\bf e}).

To characterize the volatility of the BGP updates we analyze the distribution of the number of BGP
updates received by the monitor in $1$ minute intervals. Figure~$3${\bf a} confirms the
volatile nature of BGP updates. We find that all monitors are characterized by similar
distributions $P(Z)$. Although the average number of BGP updates received per minute is quite small
($\overline{Z}_{NTT} = 250$), the peak values may occasionally exceed $10^{5}$ BGP updates per
minute. The distributions of the number of BGP updates, $P(Z)$, are positively skewed (measured
skewness values are: $\gamma_{1}($AT\&T$) = 45.7$, $\gamma_{1}(IIJ) =121.2$, $\gamma_{1}(Tinet) =
49.1$, $\gamma_{1}(NTT) = 69.1$) and the distribution tails scale as a power-law, $P(Z)\sim
Z^{-\mu}$ with $\mu=2.51\pm 0.11$ ($p=0.992$ for $IIJ$, see Appendix~\ref{sec:app:fitting} for details). We
also note that the observed power-law behavior of the tail of $P(Z)$ seems to be independent of the
aggregation window size (Fig.~$3${\bf b}).
\begin{figure*}
        \includegraphics[width = 16cm]{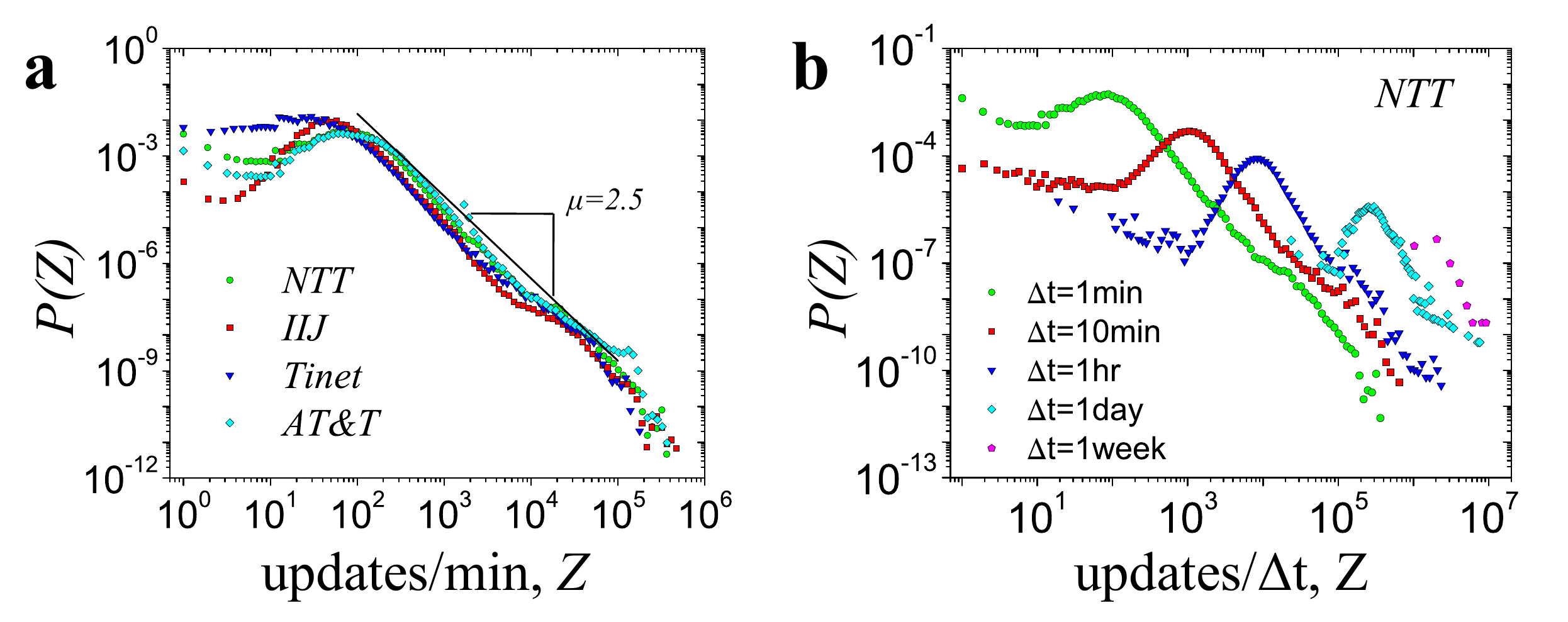}
  \caption{{\bf Extreme events in BGP dynamics.}{\bf a}, The distribution of the number of BGP
updates received by the $4$ monitors in $1$ minute interval,  All monitors collapse onto a single
master curve. Power law regression fit yields a slope of $\mu=2.3$
  {\bf b}, The distribution of number of updates $P(z)$ received by the $NTT$  monitor calculated for
  aggregation window sizes $\Delta t = 1min$, $\Delta t = 10min$, $\Delta t = 1hour$, $\Delta t = 1day$ and $\Delta t =
1week$.}
  \label{fig:fig2}
\end{figure*}
\begin{figure*}
     \includegraphics[width = 9cm]{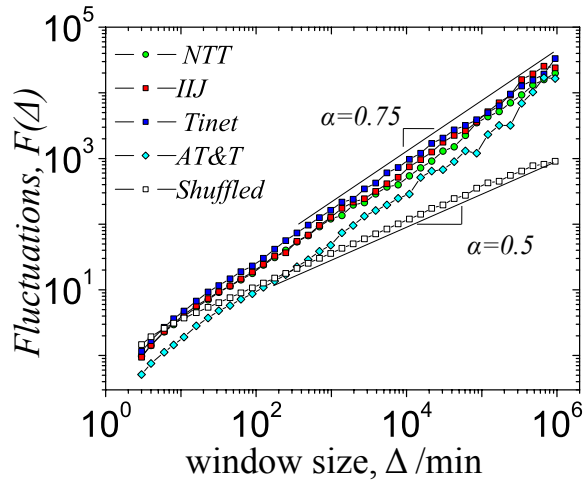}
  \caption{{\bf Correlations in the BGP update times series.} Fluctuations of the detrended BGP
update time series as a function of window size.}
\end{figure*}
The power-law distribution of the number of BGP updates implies that BGP routers should be able to
cope with surges in the number of updates exceeding the corresponding average levels by several
orders of magnitude. To understand how and when these surges occur we analyze correlation patterns
of the BGP updates. We employ three standard methods traditionally used in the time-series
analysis: auto-correlation function (ACF), power spectrum (PS), and the linear detrended fluctuation
analysis (DFA$1$) (see Methods, Apendix~\ref{sec:app:powerspectrum}, and Ref.~\cite{peng1994mosaic} for details).
\begin{figure*}
\includegraphics[width=18cm]{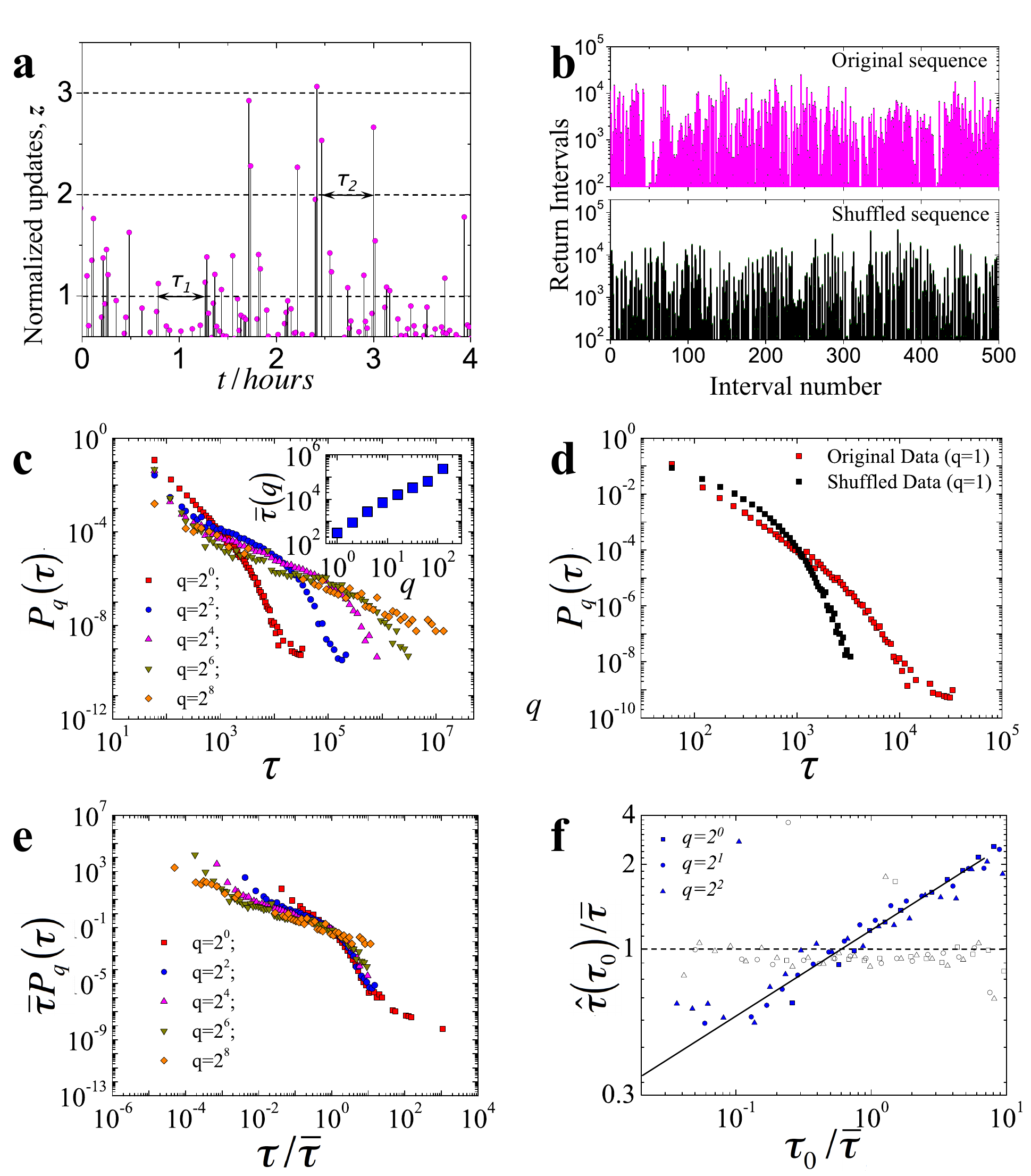}
  \caption{{\bf Return interval statistics of BGP updates.} {\bf a}, Schematic illustration of
the BGP update return intervals. Shown are the intervals $\tau_{1}$ and $\tau_{2}$ calculated for
threshold $q=1$ and $q=2$ respectively.  {\bf
  b}, Typical sequence of $500$ BGP update return intervals for $NTT$, where  $q=4$, calculated for (magenta) original
   and (black) shuffled data. {\bf c}, The distribution function $P_{q}(\tau)$ of
  BGP update return intervals of the $NTT$, calculated for different values of
  $q$. The inset depicts the average return interval $\overline{\tau}$ as a function of threshold $q$.
  {\bf d}, $P_{q}(\tau)$ for BGP update return intervals of the $NTT$ monitor calculated for $q=1$.
  Original data is shown with red while shuffled data is shown with black. {\bf e}, Scaled
  plots of the BGP return intervals for the $NTT$ monitor. {\bf f}, The mean conditional
  return interval $\hat{\tau}$  as a function of preceding return interval
  ${\tau_{0}}$ for the $NTT$ monitor. Both $\hat{\tau}$ and   ${\tau_{0}}$  are normalized with the mean return interval
  ($\overline{\tau}$).  For BGP updates without memory we expect
  $\hat{\tau}(\tau_{0})=1$, as supported by the open symbols obtained for shuffled return
  interval data.
}
 \label{fig:memory}
\end{figure*}

Even though most of BGP update events last less then $1$ minute, the duration of some of them may
exceed several minutes~\cite{labovitz2000delayed}. Thus, to avoid possible correlations associated
with long BGP updates in our subsequent analysis we use larger aggregation window size of
$\Delta(t) = 10 min$. Further, to eliminate possible spurious effects and correlations attributed
to periodic activities we also normalize the BGP update data with both intra-week and intra-day
patterns:
\begin{equation}
z(t) \equiv {Z(t) \over \overline{Z}_{w}\left(t_{week}(t)\right)
\overline{Z}_{d}\left(t_{day}(t)\right)},
\end{equation}

All three methods indicate the presence of
long-range correlations in the BGP update time-series~(see Fig.~$4$ and Fig.~\ref{fig:corr}). Specifically, we find that DFA$1$ performed for $NTT$, $IIJ$ and $Tinet$ and $AT\&T$
indicates that fluctuations grow as a power-law with aggregation window size $\Delta$,
$F(\Delta)\sim \Delta^{\alpha}$, where $\alpha = 0.75$ (Fig.~$4$). To highlight
the effects of long-range correlations in the BGP updates time series we also performed DFA$1$ for the
randomized counterparts of the BGP updates~(see Methods). In the randomized case we obtained
$F_{random}( \Delta)\sim \Delta^{\alpha}$ with $\alpha=0.5$,
which corresponds to the uncorrelated time series~(Fig.~$4$). Similar results
are obtained by ACF and PS analysis. The autocorrelation function of the BGP updates decays as a
power law over several orders of magnitude for all monitors, $ACF(\Delta z) \sim
z^{-\gamma}$~(Fig.~\ref{fig:corr}{\bf a}). We obtain similar $\gamma$  values for three monitors: $\gamma=0.5$
for $NTT$, and $\gamma=0.4$ for $IIJ$ and $Tinet$ monitors. The power spectrum density, $S(f)$,
also decays as a power-law with frequency, $S(f)\sim f^{-\beta}$, where $\beta=0.6$ for all
monitors (Fig.~\ref{fig:corr}{\bf b}). We note that the obtained values of correlation exponents
approximately conform with expected relations, $\gamma = 1-\beta$, $\alpha = {\beta + 1 \over 2}$,
and $\gamma=2(1-\alpha)$~\cite{beran1994statistics,PhysRevLett.81.729,barabasi_surface_growth,kantz2004nonlinear}.

The appearance of long-range correlations in BGP update time series indicates that  at a given time
the state of a particular BGP router is determined by its previous states. Consequently, long-range
correlations may imply the presence of memory effects in the inter-domain Internet routing. To
probe for the latter we ask, what is a typical time interval $\tau$ separating two large events.
Formally, we define a return interval $\tau(q)$ as a time separation between two consecutive events
$z(t_{1})$ and $z(t_{2})$, such that $z(t_1) > q$ and $z(t_2) > q$ (see Fig.~$5${\bf
a}). The evidence of memory in BGP update time series is seen in Fig.~$5${\bf b},
which  displays typical sequence of $500$ consecutive return intervals for the $NTT$ monitor.  The
original return interval data (shown in magenta) is characterized by ''patches'' of extreme return
intervals, while there is no such ''patches'' in the shuffled data (shown in black) obtained by
randomizing the time-order of the original series of BGP updates.

To further explore memory effects we analyze the distribution of return intervals $P_q{(\tau)}$
for the $NTT$ monitor~(Fig.~$5${\bf c}). We note that $P_q{(\tau)}$ decays slower than
the Poisson distribution, which is expected for uncorrelated data (Fig.~$5${\bf d}).
As $q$ increases, the decay of $P_q{(\tau)}$ becomes slower and the average return interval
$\overline{\tau}(q)$ increases implying that the larger events become increasingly rare (see the
inset of Fig.~$5${\bf c})). We also note that, independent of $q$, all the distributions~$P_q{(\tau)}$,
upon proper rescaling, collapse to a single master curve:
\begin{equation}
P_q{(\tau)}={1 \over \bar \tau} f \left({\tau \over \bar \tau}\right), \label{eq:scaling}
\end{equation}
where $f(x)$ does not depend on the threshold value~(see Fig.~$5${\bf e} and
Fig.~\ref{fig:SI_memory}). The resulting master curve $f(x)$ fits a stretched exponential $exp\left(-x^{-\gamma}
\right)$ with exponent $\gamma=0.5$ ($p=0.08$, see Appendix~\ref{sec:app:fitting}), which approximately
matches the observed autocorrelation exponent $\gamma=0.5$~\cite{PhysRevE.75.011128}. We note that
the observed scaling of $P_q{(\tau)}$ holds not only for $NTT$ but also for the other three
analyzed monitors (see Appendix~\ref{sec:app:return} and Fig.~\ref{fig:SI_rtn}).

Finally, to test memory effects  directly we measure the average return interval
$\hat{\tau}$ following immediately after return intervals of fixed duration $\tau_0$.
Figures~$5${\bf f} and Fig.~\ref{fig:SI_memory}. depict $\hat{\tau}$ as a function of $\tau_0$ for three
possible values of threshold $q$ (filled symbols). We observe that $\hat{\tau}$ increases as a
function of $\tau_0$, indicating that on the average longer (shorter) return intervals tend to
follow longer (shorter) intervals. In contrast, $\hat{\tau}$  is independent of preceding return
interval $\tau_0$ for randomized data (open symbols in Fig.~$5${\bf f} and Fig.~\ref{fig:SI_memory}).

\section{Discussion}
\label{sec:discussion}

In this work, we investigated the statistical properties of BGP updates. Complementing previous
studies, we confirmed that the rate of BGP updates is highly volatile, with extreme events at times
exceeding the average rates by up to 4 orders of magnitude. We established that the distribution in
the number of BGP updates received by a BGP monitor in a given time window is characterized by a
power-law tail with exponent $\mu = 2.5$. We also found (using three independent methods) that the
BGP update time series exhibit long-range correlations. The analysis of the return interval data
revealed the universal scaling in the distribution of return intervals $P_q{(\tau)}$. We also found
memory effects in the return interval data. Small (or large) return intervals separating BGP update
events are more likely to be followed by small (or large) intervals.

The observed volatility and correlation properties of the BGP update dynamics place interdomain
Internet routing into the same class of phenomena as earthquakes~\cite{lennartz08,bunde2002science}, climate~\cite{eichner03}, stock
markets~\cite{yamasaki05,preis2010complex,preis2012quantifying} and languages~\cite{gao2012culturomics,petersen2012languages}. Unlike these systems, however, the Internet routing
is a fully engineered system. The observed dynamical similarities between these stochastic systems
imply that the key mechanisms underlying Internet routing are in a certain way similar to the
mechanisms governing the dynamics of stock markets or seismic movements in the Earth crust.

As with stock market price dynamics, one would wish to be able to predict BGP dynamics, or at least
extreme events in it. To this end, one could benefit from the return interval scaling. The
established scaling of $P_{q}(\tau)$ may allow one to approximate the statistics of return
intervals for large events (characterized by large $q$ values) using the much richer statistics of
return intervals of smaller events.

The observed long-range correlations and memory effects indicate that the communication patterns
between BGP routers are an outcome of an interplay between certain semi-deterministic processes.
 Such processes are well known at the low level of the operation of an individual BGP
router (e.g. BGP route selection process). Yet this knowledge is as helpful as the knowledge about the
dynamical properties of an individual molecule in a gas---when studying the properties of this gas
(or the Internet in our case), some molecular details do matter, but most details are irrelevant.

Therefore the identification of a proper level of abstraction in modeling the dynamics of BGP
routing is an important problem for understanding Internet dynamics. The statistical analysis of
the BGP update time series that we have conducted here should serve as a basis for
validation of existing models and for developing better ones.

\label{app:a}
\section{Materials and Methods}

{\bf Intraday and Intraweek Patterns}

Consider series $Z(t)$, where $Z$ is the number of events taking place at time $t$, and $t$ is
specified as UNIX timestamps. We first define functions $t_{day}(t)$ and $t_{week}(t)$ which map
Unix timestamps $t$ to respectively specific time of the day or specific time of the week ($t_{day}
\in \left[0\colon00,~24\colon00 \right]$, $t_{week} \in
\left[Sunday,~0\colon00,~Saturday,~24\colon00 \right]$). Both $t_{day}$ and $t_{week}$ are
calculated corresponding to the GMT time zone.

The intra-day pattern, $\overline{Z}_{d}$, is then defined as the
 number of events taking place at a specific time of the day, $t_{day}$, averaged throughout the observation period:
\begin{equation}
\overline{Z}_{d}(t_{day}) = {1 \over N_{d}}\sum_{i=1}^{N_{d}} Z^{i}(t_{day}), \label{eq:id}
\end{equation}
where $N_{d}$ is the total number of days in the observation period, and  $Z^{i}(t_{day})$   is the
number of events at day $i$ at $t_{day}$.  The intra-week pattern $\overline{Z}_{w}(t_{week})$ is
defined in a similar way after first normalizing the time series with the intra-day pattern.
\begin{eqnarray}
\tilde{Z}(t) &\equiv& {Z(t) \over \overline{Z}_{d}\left(t_{day}\left(t\right)\right)},\\
\overline{Z}_{w}(t_{week}) &=& {1 \over N_{w}}\sum_{i=1}^{N_{w}} \tilde{Z}^{i}(t_{week})
\label{eq:id2}.
\end{eqnarray}
Here $N_{w}$ is the number of weeks in the observational period and $\tilde{Z}^{i}(t_{week})$ is
the normalized number of events at week $i$ at time of the week
 $t_{week}$.

The standard deviations of the intraday and intraweek patterns are defined as
\begin{eqnarray}
\sigma_{d}(t_{day}) &\equiv& \sqrt{{1 \over N_{d}}\sum_{i=1}^{N_d} \left(Z^{i}(t_{day}) -
\overline{Z}_{d}(t_{day})\right)^{2}},\\
\sigma_{d}(t_{week}) &\equiv& \sqrt{{1 \over N_{w}}\sum_{i=1}^{N_w} \left(\tilde{Z}^{i}(t_{week}) -
\overline{Z}_{w}(t_{week})\right)^{2}}
\end{eqnarray}

{\bf  Detrended Fluctuation Analysis}

Detrended Fluctuation Analysis is a method designed to study correlations in time
series ~\cite{peng1994mosaic}. Here we employ the linear version of the DFA, defined as follows.
We first calculate the cumulative BGP update time series:
\begin{equation}
y(t) = \sum_{t'=t_i}^{t} \left(z(t') - \overline{z} \right),
\end{equation}
where $t_i$ is the initial time value in the series, $z(t)$ is the original time series and
$\overline{z}$ is its average value. The cumulative time series $y(t)$ is then divided into boxes
of equal size $\Delta$. In each box, a least squares linear fit to the $y(t)$ data is performed,
representing the trend in that box. That is, for each box $\Delta$ we determine linear
approximation for the corresponding piece of the time series:
\begin{equation}
y_{\Delta}(t) = m_{\Delta}t+b_{\Delta},
\end{equation}
where $m_{\Delta}$ and $b_{\Delta}$ are the slope and the intercept of the straight line. Next we
detrend the integrated time series, $y(t)$, by subtracting the local trend, $y_{\Delta}(t)$, in
each box. The root-mean-square fluctuation of this integrated and detrended time series is
calculated:
\begin{equation}
F(\Delta) = \sqrt{{1 \over N}\sum_{t=t_{i}}^{t_f} \left[ y(t) - y_{\Delta}(t)\right]^{2}},
\end{equation}
where $N$ is the total number of points in the original time series, $t_{i}$ and $t_{f}$ are
respectively the initial and final time values in the series.

 This fluctuation measurement process is repeated at a range of different box sizes $\Delta$.
 The fluctuations typically exhibit a power law scaling as a function of box size:
\begin{equation}
F(\Delta) \sim \Delta^{\alpha},
\end{equation}
depending on the observed exponent $\alpha$ one can distinguish anti-correlated fluctuations
($\alpha < 1/2$), uncorrelated fluctuations ($\alpha = 1/2$), and correlated fluctuations ($\alpha
> 1/2$).

{\bf Data Randomization}

 To assess the significance of correlations and memory effects  in the BGP
update time series we compare original results to those obtained for randomized (shuffled)
datasets. In all experiments the randomization is performed at the most basic level: for a given
time series $Z(t)$ we obtain its randomized (shuffled) counterpart by randomly rearranging time
stamps attributed to each element in the series. Shuffled data is subsequently normalized and
binned using the same procedures as those applied to original data.

\section{Acknowledgements}
We thank H. E. Stanley, kc claffy, and D. Rybski for useful discussions and suggestions.

\section{Author Contribution} M.K., A.E., S.H., and D.K. designed research; M.K. and A.E. performed
research; M.K., and A.E. analyzed data and performed simulations;  M.K., A.E., and D.K. wrote the
manuscript; all authors discussed the results and reviewed the manuscript. M.K. and A.E.
contributed equally to this work.

\appendix

\section{The collection and pre-processing of the BGP update data}
\label{sec:app:bgp}

Our analysis is based on BGP update traces collected by the RouteViews project~\cite{routeviews}.
Routeviews collects the BGP update data from select ASes. BGP routers in these ASes are referred to
as {\em monitors}.  Monitors send  BGP updates to the Routeviews collector every time there is a
routing change. BGP updates are recorded by the Routeviews with a granularity of one second. We
focus on update traces from monitors at large transit networks in the core of the Internet.
Specifically, we analyze the BGP update time series from four monitors: $AT\&T$, $NTT$, $IIJ$, and $Tinet$.

 $AT\&T$ (American Telephone \& Telegraph) is an American multinational telecommunications corporation, headquartered in Dallas, TX. $AT\&T$ is one of the largest providers of telephone services in the United States.  $AT\&T$ also provides broadband subscription to television services.
$NTT$ (Nippon Telegraph and Telephone) is a Japanese telecommunications company headquartered in Tokyo, Japan and  is one of the largest telecommunications companies in the world in terms of revenue. $IIJ$ (Internet Initiative Japan) is the first Japan's Internet provider, which is headquartered in Tokyo, Japan. $IIJ$ is currently known as total solutions provider, offering network services, value-added outsourcing services, and cloud computing, WAN services and systems integration services. $Tinet$ (The Tiscali International Network) is an Italian Internet service provider headquartered in  Cagliari, Italy. $AT\&T$, $NTT$, and $Tinet$ are present worldwide. $IIJ$ has very strong presence in Japan and also operates in the US, UK, Germany, China, Hong Kong, Indonesia, Singapore, and Thailand.

$AT\&T$, $NTT$, $IIJ$, and $Tinet$ monitors correspond to the largest Internet Service Providers and belong to the {\em Default
Free Zone}. In other words, they have a route to every destination prefix on the Internet. Thus,
corresponding BGP update traffic is a reflection of BGP dynamics taking place in the core of the
Internet, where the BGP update volatility is believed to reach maximum rates. Our data spans $8.5$
years from mid-$2003$ through the end of $2011$.

\bigskip

{\bf Data pre-processing}
\begin{itemize}
\item
During the period of observation some of the IP addresses of the monitors changed. We identified
these changes and concatenated corresponding update time series.

\item
In the case BGP session between the monitor and the Routeviews collector is broken and
re-established, the monitor re-announces all its known paths to the collector producing a burst in
the number of BGP updates. These local artifacts of the RouteViews measurement infrastructure are
known as {\em session resets}. Session resets do not represent genuine BGP routing dynamics. We
identified and removed BGP updates corresponding to session resets using the method developed in a
course of our previous work~\cite{filterresets}.

\item
In addition to session resets we have identified and removed BGP updates corresponding to periods
of non-stationarity in the dynamics of the monitors, caused by misconfigurations and
monitor-specific events.  We refer the reader to Ref.~\cite{elmokashfi2013revisiting} for a
detailed discussion of these non-stationary periods and the methods used for their identification.

\end{itemize}

The BGP update datasets used in this work are publicly available at the Figshare repository: \url{http://figshare.com/articles/Correlation_in_global_routing_dynamics/1549778}.

\section{Correlations in BGP Updates Time Series}
\label{sec:app:powerspectrum} In this section we discuss  methods we use to characterize correlations in the BGP
update times series: Auto-Correlation Functions (ACF) and Power Spectrum (PS). Linear Detrended
Fluctuation Analysis~(DFA$1$) is discussed in the Methods section of the main text.

\begin{figure}
    \includegraphics[width = 16cm,angle=0]{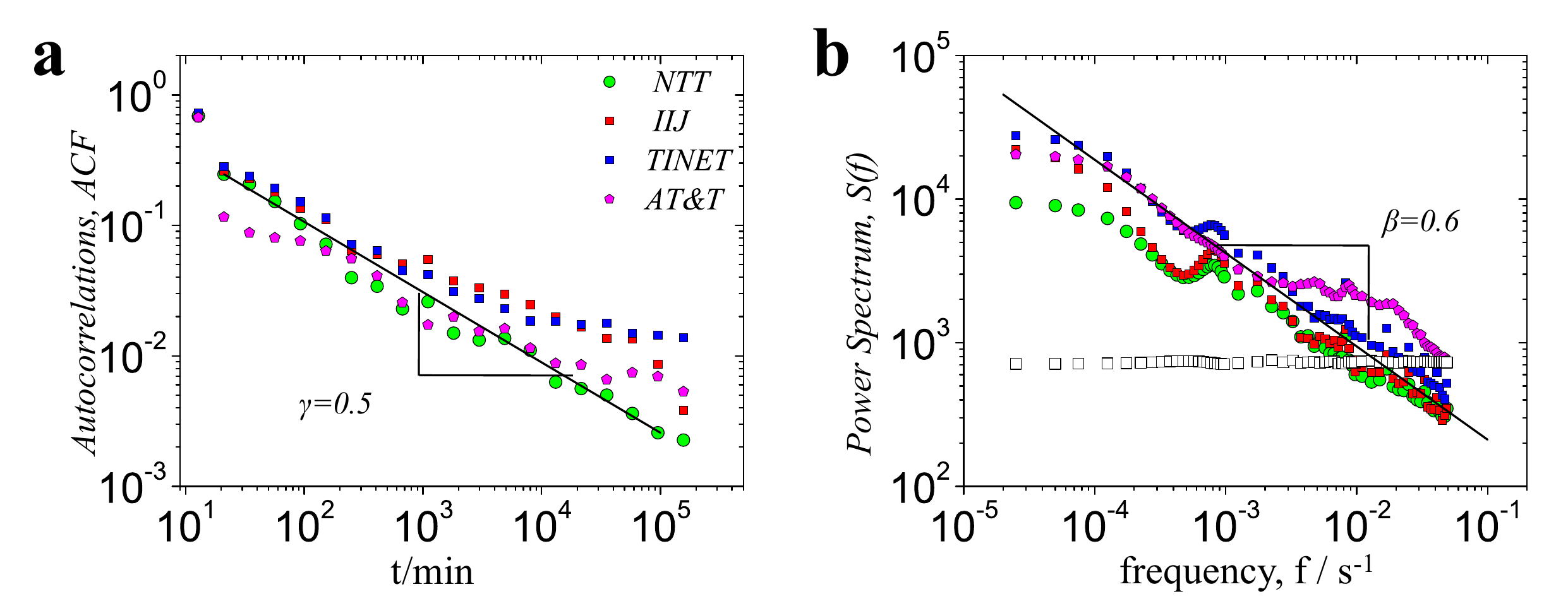} \\
  \caption{\footnotesize Correlations in the BGP update times series. {\bf a}, The autocorrelation function, ACF and  {\bf b}, The Power Spectrum S(f)}
 \label{fig:corr}
\end{figure}

\begin{figure}
     \includegraphics[width = 15cm,angle=0]{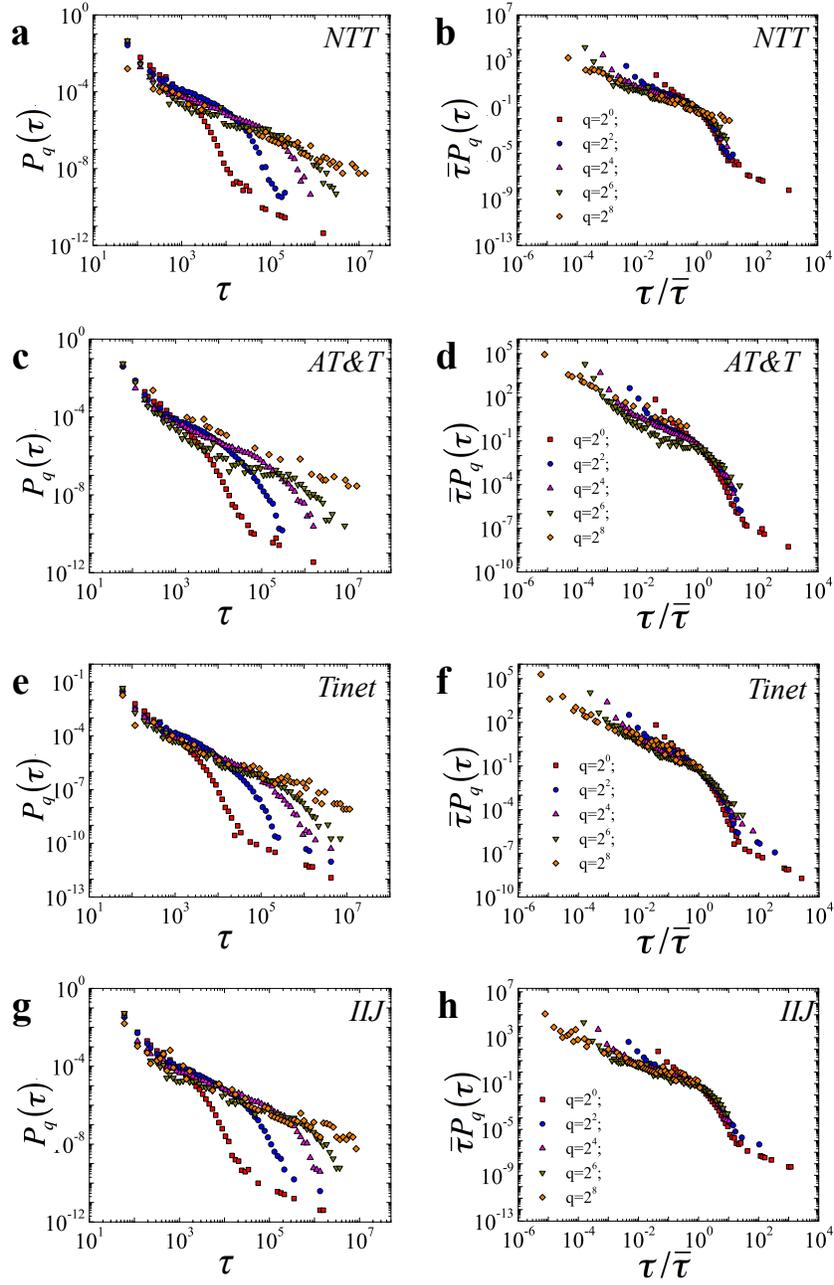} \\
  \caption{\footnotesize (Left column) The distribution of return intervals, $P_{q}(\tau)$, for BGP updates of {\bf a} $NTT$, {\bf c}
  $AT\&T$, {\bf e} $Tinet$, and {\bf g} $IIJ$ monitors. the distributions are calculated for different values of threshold $q$.
  (Right column)  Rescaled  plots of the BGP return intervals for of {\bf b} $NTT$, {\bf d}
  $AT\&T$, {\bf f} $Tinet$, and {\bf h} $IIJ$ monitors.}
  \label{fig:SI_rtn}
\end{figure}

\begin{figure}
     \includegraphics[width = 16cm,angle=0]{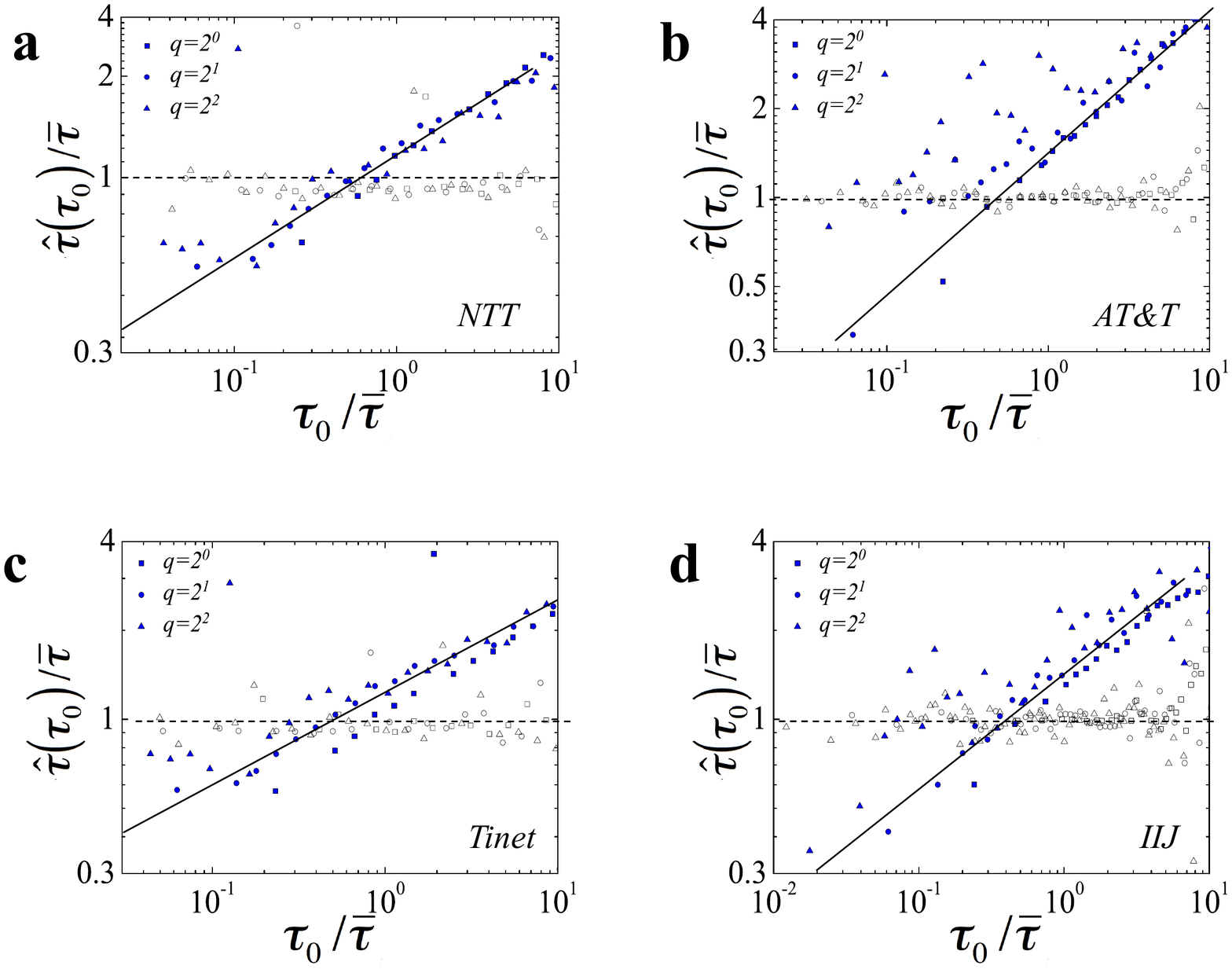}
  \caption{ \footnotesize The mean conditional  interval $\hat{\tau}(\tau_{0})$ divided by $\overline{\tau}$ as a function of ${\tau_{0}
  \over \overline{\tau}}$ for {\bf a} $NTT$, {\bf b} $AT\&T$, {\bf c} $Tinet$, and {\bf d} $IIJ$ monitors. In
time series without memory,
  $\hat{\tau}(\tau_{0})=1$, supported by the open symbols that show the shuffled return
  interval data.}
  \label{fig:SI_memory}
\end{figure}

\subsection{Auto-Correlation Function}
Correlations in the discrete time series $z(t)$ can be quantified by the
ACF~\cite{beran1994statistics}, which is defined as
\begin{equation}
ACF(\tau) ={1 \over \sigma^{2}} \sum_{t=1}^{N-\tau}(z(t)-\mu)(z(t+\tau)-\mu) \label{eq:auto},
\end{equation}
where $N$ is the total number of elements in the time series, $\mu$ and $\sigma$ are respectively
the mean and the standard deviation of the time series:
\begin{eqnarray}
\mu = {1 \over N} \sum_{t=1}^{N} z(t), \\
\sigma^{2} = {1 \over N} \sum_{t=1}^{N} \left( z(t) - \mu\right)^{2}
\end{eqnarray}
The values of the ACF range from $+1$ (very high positive correlation) $-1$ (very high negative
correlation). In the case there is no correlations, the $ACF(\tau) \approx 0$. In the case $z(t)$
is correlated for up to $t_{0}$ time steps, corresponding $ACF(\tau)$ is positive for $\tau <
t_{0}$. Among correlated time series one often distinguishes short-range and long-range
correlations. In the case of short-range correlations, $ACF$ exhibits a fast, typically
exponential, decay to zero:
\begin{equation}
ACF(\tau) \sim e^{-\tau / \zeta},
\end{equation}
where $\zeta$ is the effective correlation length. In the case long-range correlations are present,
the $ACF$ decreases to zero much slower, typically as a power-law:
\begin{equation}
ACF(\tau) \sim \tau^{-\gamma}.
\end{equation}
Figure~\ref{fig:corr}{\bf a} depicts the ACF measured for the BGP update time series. All four
monitors exhibit power-law scaling of the ACF with $\gamma = 0.5$ of the $NTT$ monitor and $\gamma =
0.4$ for other monitors. Noise in the ACF plots is caused by irregular fluctuations in the time
series as well as possible non-stationarities.

\subsection{Power-Spectrum}

Power spectrum is defined as the Fourier Transform of the ACF~\cite{beran1994statistics}:
\begin{equation}
S(f) =\int^\infty_{-\infty} ACF(\tau)e^{-2{\pi}if\tau}\,d\tau \label{eq:spec}
\end{equation}
In the case of long-range correlations characterized by $ACF \sim \tau^{-\gamma}$ , $S(f)$ also
decays as a power-law:
\begin{equation}
S(f) \sim f^{-\beta},
\end{equation}
where
\begin{equation}
\beta = 1-\alpha \label{eq:beta_vs_gamma}
\end{equation}
%a
Even though the Power-Spectrum is fully derived from the ACF, the use of the former in series
analysis often helps to decrease the noise. As seen from Fig.~\ref{fig:corr}{\bf b}, $S(f) \sim
f^{-\beta}$ for all four monitors with $\beta \approx 0.6$, which is consistent with
Eq.~(\ref{eq:beta_vs_gamma}).

Figures~\ref{fig:corr}{\bf a} and \ref{fig:corr}{\bf b} complement the results of  DFA$1$, which we report in the main text.

We note the long-range correlation exponents measured with the three methods (ACF ($\gamma$),  DFAS$1$
($\alpha$) and PSD ($\beta$) conform with expected
relations~\cite{beran1994statistics,PhysRevLett.81.729,barabasi_surface_growth}:
\begin{eqnarray}
\gamma &=& 2-2\alpha, \label{eq:alpha}\\
\alpha &=&\frac {\beta+1} 2, \label{eq:beta}\\
\gamma &=& 1-\beta \label{eq:gamma}
\end{eqnarray}

\section{Return Intervals and Memory in BGP updates}\label{sec:app:return}

Figure~\ref{fig:SI_rtn}{\bf a, c, e, g} displays the distribution of return
intervals, $P_{q}(\tau)$ calculated for all four monitors. Return intervals calculated for
different values of the threshold $q$. Upon rescaling $P_{q}(\tau)$ follow  the same master-curve
$f(x)$:
\begin{equation}
P_q{(\tau)}={1 \over \bar \tau} f \left({\tau \over \bar \tau}\right),
\end{equation}
where $f(x)$ does not depend on the threshold value and fits a stretched exponential
$exp\left(-x^\gamma \right)$, where $\gamma \approx 0.5$~(Fig.~\ref{fig:SI_rtn}{\bf b,d,f,g}). Appendix~\ref{sec:app:fitting} for details on data fitting.

We also note that all four monitors exhibit memory effects. As seen, from Fig.~\ref{fig:SI_memory},
large (small) return intervals are likely to be followed by large (small) return intervals.

\section{Statistical Tests of Data Fitting}\label{sec:app:fitting}

\begin{figure}
     \includegraphics[width = 16cm,angle=0]{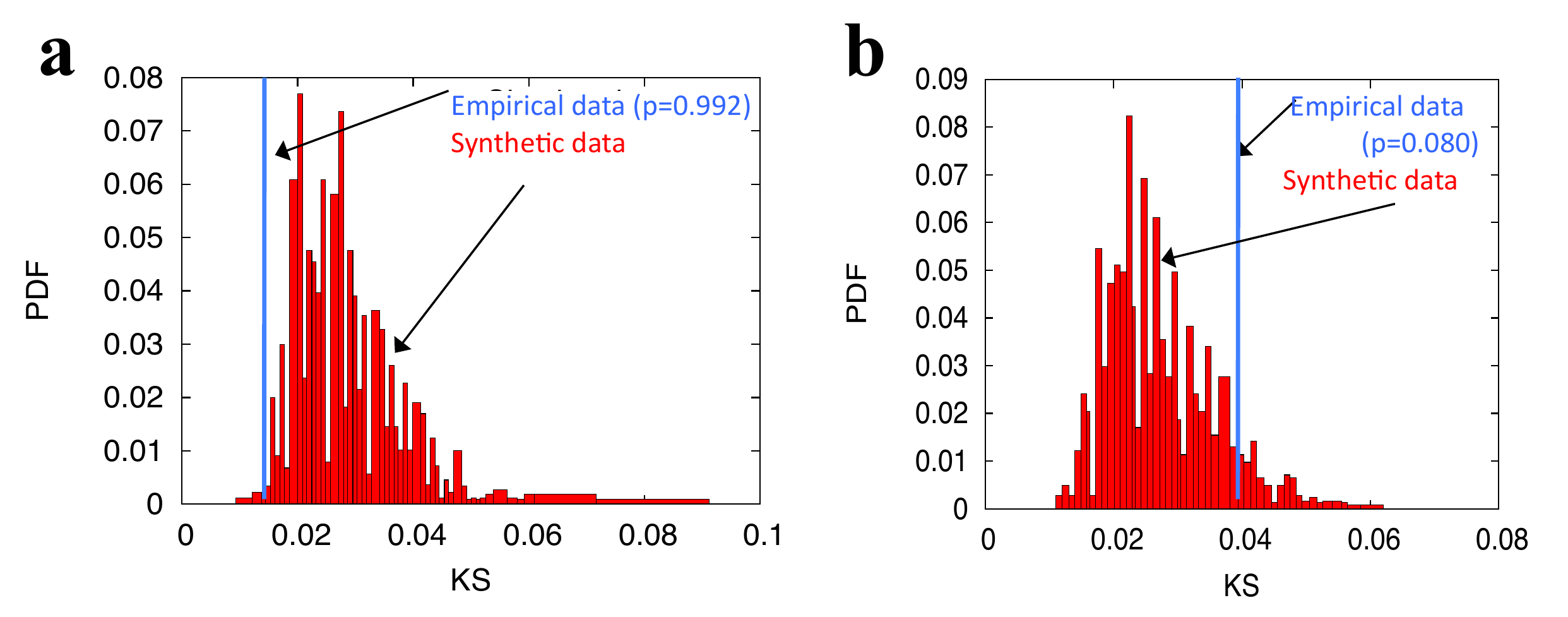}
  \caption{\footnotesize KS goodness of fit tests for {\bf a}, the distribution of number of BGP updates, $P(z)$ for the $NTT$ monitor,
  and {\bf b}, the distribution of return intervals $P_{q}(\tau)$ for the $NTT$ monitor.}
  \label{fig:data_fitting}
\end{figure}

In this section we present the results on data fitting for the distribution of
number of updates, $P(z)$, and the distribution of return intervals, $P_{q}(\tau)$, calculated for
the $NTT$ monitor. For $P(z)$, we fit time series with aggregation window size of 1 min, and for
$P_{q}(\tau)$ we fit return interval time series with $q \in (1,2,4,8,16,32)$. Note that we pick
these values since
 they yield a reasonable sample size for performing the fitting.

We tested the goodness of fit using the Kolmogorov-Smirnov (KS) goodness of fit
test~\cite{Kelton:1999}.

The KS test can be summarized as follows. Given two cumulative distribution functions (CDF),
$F_{1}(x)$ and $F_{2}(x)$ one defines the KS statistic as
\begin{equation}
KS(F_{1},F_{2}) = {\rm sup}_{x}|F_{1}(x)- F_{2}(x)|.
\end{equation}
The KS statistic can be viewed as a separation between the two distribution. The smaller the KS the
more similar the two distributions are. One starts with the calculation of the KS statistic for the
empirical Cumulative distribution function (CDF) $E(x)$ and the CDF of its best-fit, $BF(x)$, which
is referred to as $KS_{eb}$:
\begin{equation}
KS_{eb} \equiv KS(E(x),BF(x))
\end{equation}
 Then, a large number of data samples $N_{s}$ is generated using $BF(x)$, each data sample containing a
large number of values. For each data sample $i$ one calculates CDF $E_{i}(x)$ and corresponding KS
statistic
\begin{equation}
KS_{i} \equiv KS(E_{i}(x),BF(x))
\end{equation}
The KS test compares the empirical KS statistic $KS_{eb}$ with the set of synthetic values
$KS_{i}$. Goodness of fit is quantified by the $p$-value by integrating the distribution of
synthetic KS values $P(KS)$:
\begin{equation}
p = \int_{KS_{EB}}^{\infty} P(KS) {\rm d} KS
\end{equation}
$p$-values, therefore can be interpreted as the probability that the observed data was the result
of its best fit. Large $p$-value indicates that the empirical distribution matches its best fit as
good as synthetic data generated from the fit itself, whereas a small $p$-value (typically $p <
0.01$) suggests that the empirical distribution can not be the result of its best fit.

{\bf The distribution of the number of BGP updates.} We employ the maximum-likelihood method
proposed by Clauset {\it et al}~\cite{Clauset:2009:PDE:1655787.1655789} to estimate $\mu$ and
$z_{min}$ of the $P(z)$ given by
\begin{equation}
P(z) \sim z^{-\mu},
\end{equation}
The maximum likelihood estimate yields $\alpha=2.5$ and $z_{min}\approx 160$. To assess the
goodness of fit we get $KS_{eb}\approx 0.014$ and generate $N_{s}=10^{3}$ synthetic data samples
each consisting of $10^{3}$ values (Fig.~\ref{fig:data_fitting}{\bf a}), the resulting $p$-value is $p= 0.992$.

{\bf The distribution of the return intervals obtained for the $NTT$ monitor.}
We argue that the scaled distribution of the number of BGP updates return intervals decays as a
stretched exponential distribution in the form $P_{q}(\tau) \approx e^{-x^\gamma}$. In the
following we elaborate on confirming that our empirical data is statistically consistent with the
best-fit. To find the best fitting distribution, we experimented with fitting the data using the
maximum likelihood estimation to various distributions, exponential, power-law, stretched
exponential. The latter gives the best-fit with exponent $\gamma\approx 0.5 $.

To confirm that the empirical data is statistically consistent with the best-fit, we perform the KS
goodness of fit test. We generated  $N_{s} = 10^{3}$ samples of $10^{3}$ values each from the
stretch exponential fit. We then compute the $P(KS)$ synthetic samples and compared it with
$KS_{eb}\approx 0.0395$ (Fig.~\ref{fig:data_fitting}{\bf b}). The obtained $p$-value is $p = 0.08$.

\end{document}